\documentclass[aps, prl, amsmath, floats,floatfix, twocolumn, superscriptaddress]{revtex4}

\usepackage{xcolor}
\usepackage{amssymb}
\usepackage{amsmath}
\usepackage{verbatim}
\usepackage{mathrsfs}
\usepackage{amsfonts}
\usepackage{latexsym}
\usepackage{diagbox}
\usepackage{epsfig}
\usepackage{color}
\usepackage{graphicx,subfigure}
\usepackage{units}
\usepackage[spanish,english]{babel}

\usepackage{mathtools,slashed}

\usepackage[hidelinks]{hyperref}

\newcommand{\be}{\begin{equation}}
\newcommand{\ee}{\end{equation}}
\newcommand{\bea}{\begin{eqnarray}}
\newcommand{\eea}{\end{eqnarray}}

\begin{document}

\title{Anomalies in the CMB from a cosmic bounce}

\author{Ivan Agullo}
\email{agullo@lsu.edu}
\author{Dimitrios Kranas}
\email{dkrana1@lsu.edu}
\affiliation{Department of Physics and Astronomy, Louisiana State University, Baton Rouge, LA 70803, U.S.A.
}
\author{V.~Sreenath}
\email{sreenath@nitk.edu.in}
\affiliation{Department of Physics, National Institute of Technology Karnataka, Surathkal, Mangalore 575025, India.}

\begin{abstract}
We explore a model of the early universe in which the inflationary epoch is preceded by a cosmic bounce, and argue that this scenario provides a common origin to several of the anomalous features that have been observed at large angular scales in the cosmic microwave background (CMB). More concretely, we show that a power suppression, a dipolar asymmetry, and a preference for odd-parity correlations, with amplitude and scale dependence in consonance with observations, are expected  from this scenario. The model also alleviates the tension in the lensing amplitude. These signals originate from the indirect effect that non-Gaussian correlations between CMB modes and  super-horizon  wavelengths induce in the power spectrum. We do not restrict to any specific theory, but rather derive features common to a family  bouncing models. 
\end{abstract}

\maketitle

\noindent
{\bf Introduction.} Observations  have revealed features in the CMB that are in tension with the standard model of cosmology (aka $\Lambda$CDM). The signals that have attracted more attention are: (i) Absence of two-point correlations, known as power suppression; (ii) A dipolar or hemispherical asymmetry; (iii) A preference for odd-parity correlations.  These anomalies appear only at large angular scales, and are present in data from both the WMAP \cite{2013ApJS..208...20B} and Planck \cite{Akrami:2019bkn} satellites.  
The accumulated evidence makes it difficult to attribute them  to residual systematics or foregrounds and,  consequently, their interpretation as real features in the CMB is not in dispute. 
 However, each of the observed features deviates from the predictions of the $\Lambda$CDM  model at modest significances, quantified by means of their $p$-value \cite{Ade:2013nlj}. This is  the probability of obtaining, from the $\Lambda$CDM model, a temperature map with features at least as extreme as the observed ones. The Planck team associates $p$-values $\lesssim 1\% $ to each anomaly {\it separately}  \cite{Ade:2015hxq,Akrami:2019bkn}.

These low significances open the door to two interpretations. It is possible that the $\Lambda$CDM model is complete, but we  observe an atypical portion of the background radiation.   
Or that we actually observe a typical CMB, but the primordial probability distribution contains new physics at large scales. This is  a tantalizing possibility and, as emphasized in \cite{Akrami:2018vks}, it is worth exploring  new ideas since, given a theoretical model, new analyses could increase the significance of existing signals. 

The goal of this paper is to propose a common origin for the observed anomalies.  Our ideas rest  on an extension of the so-called non-Gaussian modulation, introduced in \cite{Jeong:2012df,Lewis:2011au,Schmidt:2012ky}, and further explored in \cite{Agullo:2015aba,Adhikari:2015yya}, to account for the dipolar asymmetry. The essence of this mechanism is  that, if the primordial  distribution is not Gaussian, certain features appear with higher probability in individual realizations; i.e., their $p$-values  are larger. Our model respects homogeneity and isotropy at the fundamental level, but predicts that typical realizations look significantly more anisotropic than they would in the absence of non-Gaussianity. 

The challenge to materialize this idea has been to find a model with strong enough non-Gaussianity, but yet compatible with Planck's constraints \cite{Ade:2015ava}. 
This extension of the $\Lambda$CDM model  modifies {\it only}  the standard {\it ansatz}  of an almost-scale invariant and Gaussian  primordial spectrum of perturbations.  Although there exist several  scenarios that predict a bounce  \cite{Khoury:2001wf,Lehners:2008vx,Brandenberger:2012zb, Raveendran:2017vfx, Ijjas:2016vtq,Chamseddine:2016uef,Liu:2017puc, Shtanov:2002mb,Ashtekar:2006rx,Ashtekar:2006wn,Ashtekar:2011ni,Agullo:2016tjh,Agullo:2013dla}, we will not adhere to any specific theory, but rather focus on generic predictions.  We argue that a bounce preceding inflation can induce strong non-Gaussian correlations at scales comparable to, or larger than the horizon. Though we cannot measure these correlations directly, they produce an indirect effect in the CMB that can account for the observed anomalies. 

\noindent
{\bf The model.} We work in a spatially flat FLRW universe, and model the bounce by a scale factor that behaves as $a(t)=a_B\, (1+b \, t^2)^n$ in cosmic time $t$, where $a_B$, $n$ and $b$ are constants.  The value of  the (spacetime) Ricci scalar at the bounce is $R_B=12\, n\, b$, so this  family of bounces are parametrized  by $n$ and $R_B$---the value of $a_B$ is physically irrelevant. Different theories assign different values to $R_B$ and $n$; e.g. loop quantum cosmology \cite{Ashtekar:2011ni,Agullo:2016tjh} produces $n=1/6$ and $R_B\,$ of order one in Planck units.  
 $n=1/6$ also arises in some higher-derivative scalar-tensor theories \cite{Chamseddine:2016uef,Liu:2017puc}. 
We will consider  the ranges $n\in [1/4, 1/7]$, and $R_B\in [10^{-3},1]\,\ell^{-2}_{P\ell}$, since they include all interesting cases. It has been proven that, if the matter sector is dominated by a scalar field with an appropriate potential $V(\phi)$, an inflationary phase is an attractor of phase space trajectories after the bounce \cite{Ashtekar:2011rm,Ashtekar:2009mm,Corichi:2010zp,Gupt:2013swa}.  Hence,  the goal of the bounce in our model is not to replace inflation, but to complement it by  replacing the big bang singularity and bringing the universe to an inflationary phase. 
 
\noindent
{\bf The power spectrum.} 
Scalar perturbations start their evolution in an adiabatic vacuum in the far past, when all Fourier modes of interest are in an adiabatic regime. Their evolution across the bounce  excites some of these modes, in such a way that at the onset of inflation their quantum state  differs from  the Bunch-Davies vacuum by the presence of both excitations and non-Gaussianity. 
 We have evaluated the power spectrum for different values of $n$ and $R_B$, and the result  can be  well approximated  by three power-laws:
 
\begin{eqnarray} \label{ps}
 {\cal P}_{\cal R}(\vec{k})\, \approx \, 
 \begin{cases}
  A_s\, \left(\frac{k}{k_B}\right)^{n_s-1} \;\;\; k\, > k_B\\                  
  A_s\, \left(\frac{k}{k_B}\right)^{q} \;\;\; k_I\, <\,k\, \leq k_B\\
  A_s\, \left(\frac{k_I}{k_B}\right)^{q}\,\left(\frac{k}{k_I}\right)^2  \;\;\; k\, \leq \,k_I,
 \end{cases}
\end{eqnarray}
where $k_B\, =\, a_B\, \sqrt{R_B/6}$ and $k_I\, =\, 2\,\pi\, a_I\,\sqrt{R_I/6}$ are the characteristic scales of the problem, set by the spacetime curvature at the bounce and at the onset of inflation, respectively (we use $R_I\, =\, 5\times10^{-10}\,\ell^{-2}_{P\ell}$ \cite{Akrami:2018odb}). 
Equation (\ref{ps}) can be understood as follows. Fourier modes with $k> k_B$  are more ultraviolet than $k_B$ at the time of the bounce, and consequently they are not  amplified when they propagate across the bounce.  Their spectrum is, therefore, entirely determined by inflation. The choice of potential $V(\phi)$ is encoded in the value of $A_s$ and $n_s$.  On the other hand, modes $k_I<k \leq k_B$ are significantly affected by the bounce, and for them $\mathcal{P}_{\mathcal R}(k)$ scales as $k^{q}$. 
Our simulations show that $q$ depends on $n$, and it takes {\it negative} values, equal to  $-2$, $-1.24$, $-1.1$, $-0.7$ and $-0.5$  for $n$ equal to $1/4$, $0.21$, $1/5$, $1/6$ and $1/7$, respectively. 
These values are largely independent of $R_B$. Therefore, the leading order effect of the bounce is an enhancement of ${\cal P}_{\cal R}(\vec{k})$ at infrared scales.
Finally, modes  $k<k_I$ are so infrared that they are not affected either by the bounce nor by inflation, and for them $\mathcal{P}_{\mathcal R}(k)$ is largely suppressed, with a scale dependence given by  $k^{2}$. 
 If the bounce is responsible for the anomalies in the CMB, then $k_B$ must be of the order of the pivot scale $k_*$, which today corresponds to $0.002\, {\rm Mpc}^{-1}$. We adopt $k_B=\, k_*$, which makes the effects from the bounce appear for angular multipoles $\ell \lesssim 30$ in the CMB. This is equivalent to fixing the amount of expansion from the bounce to the end of inflation.  Concrete models may come with a justification for such a choice, as it is the case e.g.\  in loop quantum cosmology \cite{Ashtekar:2016wpi}.

\noindent
{\bf Primordial non-Gaussianity} is described by the  bispectrum $B_{\Phi}( k_1, k_2,k_3)$ (see e.g.\ \cite{Maldacena:2002vr}), whose details is conveniently encoded in the function $f_{NL}(k_1,k_2,k_3)\equiv 
 B_{\Phi}( k_1, k_2,k_3)/ [P_{\Phi}( k_1)P_{\Phi}( k_2)+1
\leftrightarrow 3+2
\leftrightarrow 3]$, where $\Phi$ is the Bardeen potential and $P_{\Phi}(k)\, =\left(\frac{3}{5}\right)^2\, \frac{2\pi^2}{k^3} \, {\cal P}_{\cal R}$. An exact calculation of $f_{NL}$ requires knowledge of the gravitational action and the matter content of the universe at the bounce. Our goal  is rather to obtain an estimation of its overall form, common to all models.  The shape of  $f_{NL}(k_1,k_2,k_3)$ can be obtained  by using Cauchy integral theorem, and by noticing that its amplitude  is dominated by the pole with smaller positive imaginary part in the third-order gravitational action (see section V in \cite{Agullo:2017eyh}). In presence of a bounce, this is the pole introduced by the global minimum of the scale factor $a(\eta)$, that in conformal time is at $\eta_p=i\, \alpha \sqrt{6/R_B}=i\, \alpha/k_B$, with $\alpha =\sqrt{\frac{n\, \pi}{2}}\,  \frac{\Gamma[1-n]}{\Gamma[3/2-n]}$, where $\Gamma[x]$ is the Gamma-function. This general argument tells us that 
\be \label{fnl} f_{NL}(k_1,k_2,k_3)\approx \mathfrak{f}_{NL}\,\times  e^{-\alpha\, (k_1+k_2+k_3)/k_B}\, , \ee
where $\mathfrak{f}_{NL}$ parameterizes our ignorance about its amplitude. We see that the scale dependence  of non-Gaussianity is controlled by $R_B$ and $n$. Concrete bouncing models may add additional features to $f_{NL}(k_1,k_2,k_3)$, such as  oscillations or other  finer details, but (\ref{fnl}) approximates well its overall shape.  
We have checked this in the concrete scenario explored in \cite{Agullo:2017eyh}. For very infrared wavenumbers $k_i$, we expect $ f_{NL}(k_1,k_2,k_3)$ to become small, for the same reason as the  ${\cal P}_{\cal R}(k)$ does. This is not captured by (\ref{fnl}), but will be incorporated in our calculation by the effective infrared cut-off that the shape of $\mathcal{P}_{\mathcal R}(k)$ introduces. As mentioned above, if $k_B$ is close to $k_*$ then the non-Gaussianity  (\ref{fnl})  is restricted to the most infrared scales in the CMB, and is large only when super-horizon modes are involved.

\noindent
{\bf Non-Gaussian modulation.}  
Super-horizon perturbations {\it can} impact the CMB {\it if} they are correlated with sub-horizon modes.  We follow ideas introduced in \cite{Jeong:2012df,Lewis:2011au,Schmidt:2012ky,Agullo:2015aba,Adhikari:2015yya} to compute the bias in the statistics of the gravitational potential $\Phi(\vec k_1)$ induced by  long wavelength modes $\Phi(\vec q)$.  
At the lowest non-vanishing perturbative order, we have
 \bea \label{ngmod} \langle \Phi_{\vec k_1}\Phi^{\star}_{\vec k_2}\rangle |_{\Phi_{\vec q}}= (2\pi)^3\, \delta(\vec k_1-\vec k_2)\, P_{\Phi}(\vec k_1)&&\nonumber \\
 +   f_{NL}(\vec k_1,-\vec k_2)\, \frac{1}{2}\, \big(P_{\Phi}(\vec k_1)+P_{\Phi}(\vec k_2)\big)\, \Phi_{\vec q}&& \, ,\eea
where $\vec q$ must take the value  $\vec q=\vec k_1-\vec k_2$. As we can see, $\Phi(\vec q)$ introduces ``non-diagonal'' terms, proportional to both the magnitude of $\Phi(\vec q)$ and  the intensity of the correlations, $f_{NL}$. These terms translate to anisotropic features in the CMB. 
 In a typical patch of the universe,  one expects $|\Phi(\vec q)|$ to be of the same order as  $\sqrt{P_{\Phi}(\vec q)}$. If, on the other hand, one averages over many patches, these contributions vanish;  as it must be, since our model respects isotropy at the fundamental level.  The non-diagonal terms in (\ref{ngmod}) induce similar contributions to the CMB temperature covariance matrix 
\be \label{abiposh} \langle a_{\ell m} a^*_{\ell' m'}\rangle= C_{\ell}\, \delta_{\ell\ell'}\delta_{m m'} +(-1)^{m'}\,\sum_{LM} A^{LM}_{\ell\ell'}\,  C^{LM}_{\ell m \ell' -m'} \nonumber \,, \ee
where  $C^{LM}_{\ell m \ell' m'}$ are Clebsch-Gordan coefficients. We have encoded the non-Gaussian modulation in  $A^{LM}_{\ell\ell'}$, known as the Bipolar Spherical Harmonic (BipoSH) coefficients (see \cite{Hajian_2003, Joshi:2009mj}). They organize the modulation in an efficient manner: $L$ and $M$ indicate the ``shape'' of the modulation, while $\ell,\ell'$ account for a posible variation of the modulation amplitude at different scales in the CMB.   The monopole,  $A^{00}_{\ell,\ell'}\propto \delta_{\ell\ell'}$, shifts the value of the spherically symmetric angular  spectrum  $C_{\ell}$, while the dipole $A^{1M}_{\ell\ell'}\propto \delta_{\ell+1,\ell'}$ introduces correlations between multipoles $\ell$ and $\ell+1$. Our model  cannot predict the exact value of the BipoSH coefficients in the sky, as they depend on a concrete realization of the mode $\Phi(\vec q)$. But we can compute their mean square values 
 \bea \label{varalpha}    \sqrt{\langle|A^{L M}_{\ell \ell' }|^2\rangle}&\approx&\, \left[\frac{1}{2\pi} \, \int dq\, q^2 \, P_{\Phi}(q) \, |\mathcal{C}_{\ell \ell'}^L(q)|^2\, \right]^{1/2} \nonumber \\ & & \times C^{L0}_{\ell 0 \ell' 0}\,  \sqrt{\frac{(2\ell+1)(2\ell'+1)}{4\pi\, (2L+1)}}\, , \eea
where we have defined
 \be \label{Ctilde}  \mathcal{C}_{\ell \ell'}^{L}(q)\equiv \frac{2}{\pi}\int dk_1k_1^2  (i)^{\ell-\ell'}\Delta_{\ell}(k_1)\Delta_{\ell'}(k_1) P_{\phi}(k_1) G_L(k_1,q) \nonumber .\ee
In this expression, $f_{NL}$ has been expanded using Legendre polynomials $P_L(\mu)$ as $f_{NL}(\vec k_1,\vec q)= \sum_L G_L(k_1,q)\, \frac{2L+1}{2}\, P_L(\mu)$, with $\mu=\vec k_1\cdot \vec q$. 
Thus, the $\mu$-dependence of $f_{NL}(\vec k_1,\, \vec q)$ translates to the $L$-dependence of the BipoSH coefficients. To derive (\ref{varalpha}) we have  used that $f_{NL}(\vec k_1,\vec q)$ is larger for $q\ll k_1$.

\noindent
{\bf Power suppression, lensing, and parity.} A lack of two-point correlations $C(\theta)\equiv  \langle \delta T(\hat n)\delta T(\hat n')\rangle$, $\cos \theta\equiv \hat n\cdot \hat n'$, for  $\theta> 60^\circ$ was noticed by COBE and WMAP \cite{Hinshaw:1996ut,Efstathiou:2003tv}, and confirmed by Planck. 
The observed value of the estimator $S_{1/2}\equiv\int_{-1}^{1/2} [C(\theta)]^2\, d(\cos\theta)$ \cite{Spergel:2003cb}, which measures the total amount of correlations in  $\theta > 60^\circ$, is $S^{\rm obs}_{1/2}\approx 1500\,  \mu K^4$ \cite{Ade:2015hxq,Akrami:2019bkn} (see  \cite{Schwarz:2015cma} for further details), instead of $S_{1/2}\approx 45000 \mu K^4$ predicted by the $\Lambda$CDM model. The $p$-value of $S^{\rm obs}_{1/2}$ is a fraction of a percent \cite{Ade:2015hxq,Akrami:2019bkn,Schwarz:2015cma}.  Our model can account for such a suppression, but with an important subtlety. 
The monopolar modulation introduced by $A^{00}_{\ell\ell}$ does not change the mean value of  $S_{1/2}$, but rather it modifies its variance, increasing the range of typical statistical ``excursion'', both to {\it larger and smaller} values away from the mean. In this sense, our model does not predict a power suppression, but rather it increases the probability of observing it.  We have computed the value of $\mathfrak{f}_{_{\rm NL}}$ that makes the probability of observing $S_{1/2}\leq S^{\rm obs}_{1/2}$ approximately equal to  $20\%$ (we have approximated the statistics of  $S_{1/2}$ by a Gaussian; corrections are higher order in non-Gaussianity), and show them in Table \ref{t:shalf}. Remarkably, these values of $\mathfrak{f}_{_{\rm NL}}$ are of the same order found in loop quantum cosmology \cite{Agullo:2017eyh}. Therefore, although the ``bare'' power spectrum (\ref{ps}) is enhanced with respect to the $\Lambda$CDM value at large angles, if the correlations with super-horizon scales are strong, the probability of observing  $S^{\rm obs}_{1/2}\approx 1500\,  \mu K^4$ is high, and the observed suppression cannot be considered anomalous.

\begin{figure}
 \centering
 \includegraphics[width=\columnwidth]{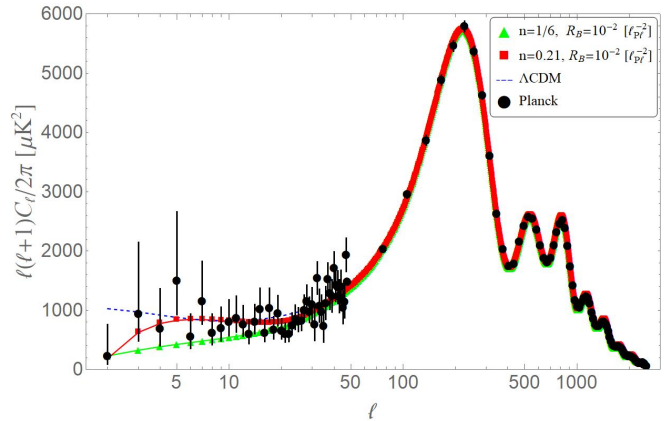}
 \caption{\label{fig:Cl} Temperature angular power spectrum for two representative bouncing models,  compared with the predictions from the $\Lambda$CDM model with the standard ansatz, and data from  Planck \cite{Akrami:2019bkn}. Both bouncing models produce $S_{1/2}= 1500 \mu K^4$.  We have used the best-fit values for $A_s$ and $n_s$.}
\end{figure}

\begin{figure}
 \centering
 \includegraphics[width=\columnwidth]{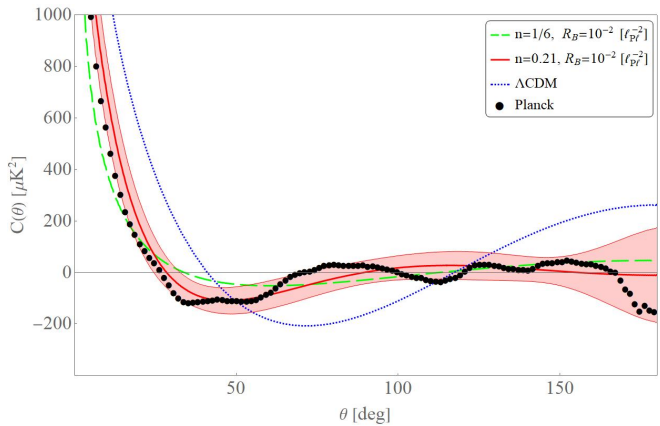}
 \caption{\label{fig:C2} Angular two-point correlations $C(\theta)$. The shadowed region is the cosmic variance of the  curve $n=0.21$, $R_B=10^{-2}\ell_{P\ell}^{-2}$, and it shows great agreement with data. The same happens for $n=1/6$. In contrast, data is clearly out of the cosmic variance region (not shown; see Fig.\ 2 in \cite{Akrami:2019bkn}) of the $\Lambda$CDM curve for   $\theta\sim 75^\circ$, and $\theta> 170^\circ$.}
\end{figure}

Next, we can  compute other effects that our model {\it predicts} must come together with a suppression. This is the goal of the rest of the paper.  First, we plot in Figures \ref{fig:Cl} and \ref{fig:C2} the details of $C_{\ell}$ and $C(\theta)$. To quantify how well our results agree with data,  we have carried out a Markov chain Monte Carlo (MCMC) analysis, using TT and low-$\ell$ $EE$ data \cite{Aghanim:2019ame}, by using the CosmoMC software \cite{Lewis:2002ah}. We have found that, although all bounces  considered here account for $S^{\rm obs}_{1/2}$ (except $n=1/4$), not all fit the details of the data equally well. Bounces for which the tilt $q$ of the power spectrum  
 is more negative, do better. For instance, for $n=0.21$ we have $q=-1.24$, and this value results in a significant improvement in $\chi^2$ of $\Delta \chi^2=-6.4$, relative to $\Lambda$CDM. Hence, this model not only reproduces the overall suppression, but it also fits  the details of $C_{\ell}$ better.  Values of $q$ closer to zero, such as $q=-0.7$ or lower, are not favored from the point of view of $\chi^2$. In this likelihood analysis we do not consider $n$, $R_B$ and $\mathfrak{f}_{_{\rm NL}}$ as free parameters; they rather must come out as predictions from individual theories.

  \begin{table}
\centering
    \begin{tabular}{|l||c| c|c|c|c|c|}
    \hline
    \hline
      \diagbox{$R_B$}{$n$} & $1/4$ & 0.21&  $1/5$& $1/6$ & $1/7$ \\
    \hline \hline
$\ \ \ \ \ 1\, \ell^{-2}_{P\ell}$ & -& 959& 1334& 3326& 5031\\ \hline
$10^{-1}\, \ell^{-2}_{P\ell}$ & -& 1560 & 2065& 4454& 6298\\ \hline
$10^{-2}\, \ell^{-2}_{P\ell}$ & -& 2573 & 3238 & 6066& 8024\\ \hline
$10^{-3}\, \ell^{-2}_{P\ell}$ & -& 4372& 5234& 8518& 10530\\
    \hline
    \hline
    \end{tabular}
\caption{\label{t:shalf}Values of the amplitude $\mathfrak{f}_{_{NL}}$ that  make the probability of observing  $S^{\rm obs}_{1/2}\leq 1500\, \mu K^4$ approximately equal to $20\%$. For $n\leq1/4$, the tilt $q$ of the power spectrum is too negative, and the non-Gaussian modulation cannot produce the observed suppression.}
\end{table}

We  find that the  suppression is accompanied by two additional effects in the $C_{\ell}$'s. On the one hand, power  suppression  induces a change in the lensing parameter $A_L$, making it closer to one than in the $\Lambda$CDM model. The relation between power suppression and the value of $A_L$ has been recently pointed out in \cite{Ashtekar:2020gec}, and we confirm it in our model. More precisely, when we include $A_L$ as a free parameter in our MCMC analysis, we obtain that the mean and standard deviation of the marginalized distribution of $A_L$ is $A_L=1.179\pm  0.092$, for $n=0.21$, $R_B=10^{-2}\ell_{P\ell}$.  Other values of $n$ and $R_B$ produce similar results. This is to be compared with the $\Lambda$CDM value, $A_L=1.243\pm  0.096$. Therefore, our model  alleviates the tension pointed out in \cite{DiValentino:2019qzk} regarding the value of $A_L$, in the sense that $A_L=1$ becomes well inside the 2-$\sigma$ region---without introducing spatial curvature, and hence avoiding the possible ``crisis in cosmology'' advocated in \cite{DiValentino:2019qzk}.

\begin{figure}
 \centering
 \includegraphics[width=\columnwidth]{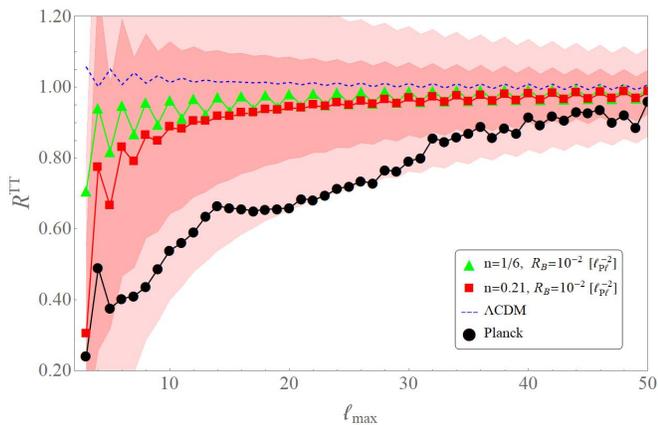}
 \caption{\label{fig:RTT}  $R^{TT}$ versus $\ell_{\rm max}$.}\end{figure}

On the other hand, we  observe that the power suppression also produces a  preference for odd-parity multipoles, as measured by $R^{TT}(\ell_{\rm max})=D_+(\ell_{\rm max})/D_-(\ell_{\rm max})$, where $D_{+,-}(\ell_{\rm max})$ is the average value of $\ell(\ell\, +\, 1)C_{\ell}/2\pi$ in even (+) or odd (-) multipoles, up to $\ell_{\rm max}$ \cite{Ade:2015hxq}.
In Fig.\ \ref{fig:RTT}  we show $R^{TT}$ versus $\ell_{\rm max}$, and the  $1\sigma$ and $2\sigma$ cosmic variance region for $n=0.21$, $R_B=10^{-2}\ell_{P\ell}$, (see also Fig.\ 25 in \cite{Akrami:2019bkn}). In contrast to $\Lambda$CDM, our model produces  a clear preference for odd multipoles  (i.e.\ $R^{TT}(\ell_{\rm max})<1$).
 
We have  also checked that, although the values of $\mathfrak{f}_{_{NL}}$ in Table \ref{t:shalf}  are significantly larger than one, the perturbative expansion remains under control, due to the smallness of ${\cal P}_{\cal R}\ll 1$.

\noindent
{\bf A dipolar modulation}
 in the CMB has been consistently observed  in data from  WMAP  \cite{Eriksen:2003db} and Planck \cite{Ade:2013kta,Ade:2015hxq,Akrami:2019bkn}. In terms of the BipoSH coefficients, this signal can be explained from a non-zero value of $A^{1M}_{\ell\,\ell+1}$. Planck's observations have been reported   in terms of  

\be \label{bb} A_1(\ell) \equiv \frac{3}{2}\sqrt{\frac{1}{3\pi} \sum_M |A^{1M}_{\ell,\ell+1} \,\times  \mathcal{G}_{\ell}^{-1}|^2 } \, ,\ee
where $\mathcal{G}_{\ell}\equiv (C_{\ell}+C_{\ell+1})\sqrt{\frac{(2\ell+1)(2\ell+2)}{12\pi}}\, \, C^{10}_{\ell,0,\ell+1,0}$ is the so-called form factor. 
The signal has been reconstructed in bins of width $\Delta \ell=64$, up to $\ell_{\rm max}=512$, and  $A_1$  deviates significantly ($\sim 3 \sigma$) from what is expected from  $\Lambda$CDM only in the first bin, were $A^{\rm obs}_1=0.068\pm 0.023$ \cite{Ade:2015hxq}.
\begin{figure}
 \centering
 \includegraphics[width=\columnwidth]{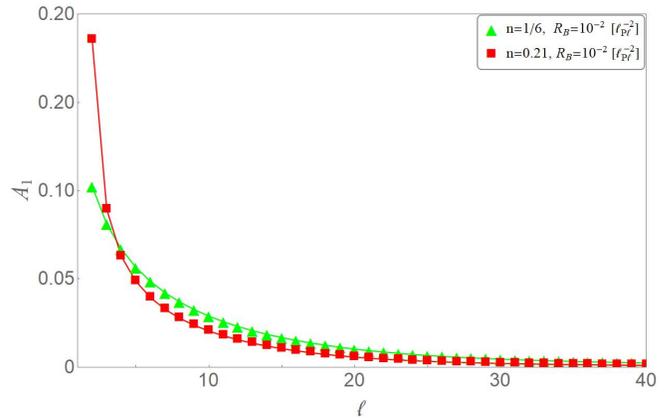}
 \caption{\label{fig:A1} Amplitude of the  dipolar modulation, as quantified by $A_1(\ell)$. The power suppression discussed above contributes to increase the amplitude of $A_1(\ell)$ at low $\ell$'s. In $\Lambda$CDM, $A_1(\ell)=0$.}
\end{figure}

Figure \ref{fig:A1} shows  our result for $A_1(\ell)$, again for  the same representative values of $n$ and $R_B$.    
We obtain that $A_1$ is large only for low multipoles $\ell \lesssim 30$. 
Although details vary slightly among different  models,  the averaged value of the amplitude for $\ell \lesssim 30$ is also in  consonance with  observations.

The Planck satellite has also looked for a quadrupolar modulation,  and found that the results are compatible with what is expected from the $\Lambda$CDM model \cite{Akrami:2018odb}.  We have computed the amplitudes of  $A^{LM}_{\ell\ell’}$ for $L>1$ in our model, and checked that they all satisfy Planck's constraints (for details, see \cite{aks}). 

\noindent
{\bf Discussion.} The  anomalies in the CMB include strong deviations from scale invariance, isotropy and parity.  It is precisely this heterogeneous character that has made  the search  for a common origin a challenging task.  We have argued that an extension of the $\Lambda$CDM model, where a cosmic bounce precedes the inflationary era, can  collectively account for these  signals, as the result of  the modulations that  very long wavelength perturbations imprint on  CMB scales.

We have not adhered to any  concrete bouncing theory, but rather introduced a series of approximations to estimate the effects of a generic bounce. The values of the parameters needed to account for the anomalies are in consonance with those coming out from concrete theories. In fact, we have complemented our analysis with calculations  using the bounce predicted by  loop quantum cosmology, and have checked that our approximations are well justified. Further detailed calculations will appear in a companion publication.

We conclude that it may be premature to disregard the large scale anomalies as mere flukes of the $\Lambda$CDM model, and  advocate the fascinating possibility that they are imprints of pre-inflationary physics,  which  carry  information about that extreme epoch. Future work will focus on extending our predictions to tensor modes, in order to construct additional ways to test our ideas. 

%%%%%%%%%%%%%%%%%%%%%%%%%%%%%%%%%%%%%%%%%%
\noindent
 {\bf Acknowledgements} 
 \acknowledgments{We specially thank Boris Bolliet for several  discussions, inputs, and initial collaboration in this project. We have  benefited from  discussions with A.  Ashtekar,  B.  Gupt, J. Olmedo, J. Pullin, and P. Singh. We thank B. Gupt for assistance with Planck data. This work is supported by the NSF CAREER grant PHY-1552603, and from the Hearne Institute for Theoretical Physics. V.S. was supported by Louisiana State University and Inter-University Centre for Astronomy and Astrophysics during different stages of this work. This research was conducted with high performance computing resources provided by Louisiana State University (http://www.hpc.lsu.edu),  and  based on observations obtained with Planck (http://www.esa.int/Planck), an ESA science mission with instruments and contributions directly funded by ESA Member States, NASA, and Canada.}

\bibliography{Refs}

\end{document}